\input epsf
\input harvmac
\noblackbox
\newcount\figno
\figno=0
\def\fig#1#2#3{
\par\begingroup\parindent=0pt\leftskip=1cm\rightskip=1cm\parindent=0pt
\baselineskip=11pt
\global\advance\figno by 1
\midinsert
\epsfxsize=#3
\centerline{\epsfbox{#2}}
\vskip 12pt
\centerline{{\bf Figure \the\figno:} #1}\par
\endinsert\endgroup\par}
\def\figlabel#1{\xdef#1{\the\figno}}

\font\cmss=cmss10
\font\cmsss=cmss10 at 7pt
\def\rlx{\relax\leavevmode}
\def\inbar{\vrule height1.5ex width.4pt depth0pt}
\def\IC{\relax\,\hbox{$\inbar\kern-.3em{\rm C}$}}
\def\IN{\relax{\rm I\kern-.18em N}}
\def\IP{\relax{\rm I\kern-.18em P}}
\def\ZZ{\rlx\leavevmode\ifmmode\mathchoice{\hbox{\cmss Z\kern-.4em Z}}
 {\hbox{\cmss Z\kern-.4em Z}}{\lower.9pt\hbox{\cmsss Z\kern-.36em Z}}
 {\lower1.2pt\hbox{\cmsss Z\kern-.36em Z}}\else{\cmss Z\kern-.4em Z}\fi}
\def\narrowplus{\kern -.04truein + \kern -.03truein}
\def\narrowminus{- \kern -.04truein}
\def\narrowminussub{\kern -.02truein - \kern -.01truein}

\def\sqr#1#2{{\vcenter{\vbox{\hrule height.#2pt
            \hbox{\vrule width.#2pt height#1pt \kern#1pt
                  \vrule width.#2pt}\hrule height.#2pt}}}}


\lref\bfss{T.~Banks, W.~Fischler, S.~Shenker and L.~Susskind,
hep-th/9610043.}

\lref\wt{W.~Taylor IV, hep-th/9611042.}

\lref\tdsd{L.~Susskind, hep-th/9611164.}

\lref\jsch{J.~ Schwarz, Phys. Lett. {\bf B367} (1996) 97.}

\lref\ganor{O.~Ganor, S.~Ramgoolam and W.~Taylor, hep-th/9611202.}

\lref\rPKT{P. K. Townsend, Phys. Lett. {\bf B350} (1995) 184.}
\lref\rWSD{E. Witten, Nucl. Phys. {\bf B443} (1995) 85.}

\lref\rschmid{C. Schmidhuber, Nucl. Phys. {\bf 467} (1996) 146.}
\lref\rtown{P. K. Townsend, Phys. Lett. {\bf B373} (1996) 68.}
\lref\rNS{N. Seiberg, unpublished.}
\lref\rBS{T. Banks and N. Seiberg, to appear; T. Banks, Lecture at the
Fourteenth Jerusalem Winter School for Theoretical Physics.}
\lref\rDHIS{ M. J. Duff, P. S. Howe, T. Inami, and K. S. Stelle, Phys. Lett.
{\bf B191} (1987) 70.}
\lref\rDL{M. J. Duff and J. X. Lu, Nucl. Phys. {\bf B390} (1993) 273.}
\lref\rM{L. Motl, hep-th/9701025.}

\Title{\vbox{\hbox{hep--th/9702101}\hbox{SU-ITP-97-6,
IASSNS--HEP--97/9}}}{\vbox{\centerline{Rotational Invariance in the
M(atrix)
Formulation}\vskip3pt\centerline{of Type IIB Theory}}}

\smallskip
\centerline{Savdeep Sethi\footnote{$^\ast$} {sethi@sns.ias.edu} }
\medskip\centerline{\it School of Natural Sciences}
\centerline{\it Institute for Advanced Study}\centerline{\it
Princeton, NJ
08540, USA}
\vskip 0.15in
\centerline{and}
\vskip 0.15in
\centerline{Leonard Susskind\footnote{$^\dagger$}
{susskind@dormouse.stanford.edu} }
\medskip\centerline{\it Department of Physics}
\centerline{\it Stanford University}\centerline{\it Stanford, CA
94305-4060,
USA}

\vskip .3in

The matrix model formulation of M-theory can be generalized by
compactification to
ten-dimensional type II string theory, formulated in the infinite
momentum
frame. Both the type IIA and IIB string theories can be formulated
in this way.
In the M-theory and type IIA
cases, the transverse  rotational invariance is manifest, but in the
IIB case, one
of the transverse dimensions materializes in a completely different
way from the
other seven.  The full O(8) rotational symmetry then follows in a
surprising
way
from the electric-magnetic duality of supersymmetric Yang-Mills
field theory.
\vskip 0.1in
\Date{2/97}


\newsec{Introduction}

 Whatever describes the fundamental degrees of freedom of string
theory must be
capable of combining into an incredible variety of different objects in
different regions of moduli space. Shenker has used the word
`protean,' which
according to our dictionary means ``readily assuming different
shapes or roles."
For example, by varying the moduli, the degrees of freedom have
to rearrange
themselves from type IIA strings to heterotic, type I, type IIB
strings, a variety
of p-branes, D-branes, and eleven-dimensional gravitons. Even more
remarkable is
the ability of the system to manufacture new space dimensions in
limits where
standard reasoning would lead one to think dimensions should
disappear. An
example of this phenomenon occurs when M-theory is compactified on
a two-dimensional torus,
and the area of the torus is shrunk to zero. Since $11-2=9$,
conventional logic
would lead to the conclusion that the theory becomes
nine-dimensional. In fact,
it becomes ten-dimensional type IIB string theory. It is clear from these
examples that the constituent objects which underly the theory must
be very unusual. Recently, evidence has accumulated that
M(atrix) theory may have exactly the
right
``protean bits" to describe this rich variety of objects. In this
note, we want
to add one more example of the `protean' nature of M(atrix) theory by
showing how the tenth  direction of type IIB theory emerges.

In the matrix formulation of M-theory \bfss,  some of the
symmetries of the
system are manifest, while others are not. For example, symmetry under
transverse
rotations of the infinite momentum frame is manifest, but symmetry under
rotations which mix the transverse and longitudinal directions is
not. When some of the transverse dimensions of M-theory are toroidally
compactified, the resulting theory can be described in terms of
type IIA string
theory \rPKT, \rWSD, \rDHIS. The transverse
rotational symmetry
in the noncompact subspace is still manifest. On the other hand,
T-duality is not
at all obvious in the nonperturbative M(atrix) theory. It was
argued in \tdsd\
and \ganor\ that T-duality of the type IIA theory is a consequence of the
electric-magnetic duality (S-duality) of 3+1-dimensional
supersymmetric Yang-Mills theory with four supersymmetries. In this
note, we
consider the matrix
formulation of type IIB theory. In this case, the transverse rotational
symmetry is
nonmanifest, and provides a nontrivial consistency test for the M(atrix)
description.  In the following discussion, we
will again invoke S-duality of the 3+1-dimensional supersymmetric
Yang-Mills
theory to prove that there is a full $O(8)$ transverse rotational
symmetry. As
a
consequence of our result, we will be able to confirm Seiberg's
prediction of
the existence of a
superconformal fixed point with $O(8)$ symmetry in strongly-coupled
2+1-dimensional Yang-Mills with eight supersymmetries.

Let us first consider the origin of the extra dimension which
appears when M-theory is compactified on a two-torus of vanishing
area. To determine how the extra dimension appears is not too
difficult. It is
fairly clear that the momentum conjugate to this
coordinate is the  conserved
wrapping number of two-branes on the two-torus \jsch. This quantum number
plays the role
of the Kaluza-Klein momentum in the new direction, and in the limit of
vanishing
area, the energy gap for this excitation vanishes and the dimension
becomes
non-compact. Obviously a consistent interpretation requires that the
physics, if
not the formalism, be invariant under the rotation of all eight
transverse
dimensions. Since M(atrix) theory can be compactified on tori, this
symmetry
requirement provides an interesting test for the theory.

\newsec{Rotational Invariance}

We begin with a word on notation. The eleven dimensions of M-theory
will be
labelled by
$(t,X^1, \ldots, X^9, X^{11})$, where $t$ is time, $(X^1, \ldots,
X^9)$ are the
transverse space
coordinates, and $X^{11}$ is the longitudinal direction of the infinite
momentum frame. To obtain type IIB string theory, we compactify two
of the
transverse coordinates on a two-torus. The torus is specified by
its complex
structure modulus, $\tau$, and its area. The complex parameter,
$\tau $, maps
to
the complexified type IIB string coupling,
$$ \tau = \chi + i e^{-\phi}, $$
where $\chi$ is the Ramond-Ramond scalar, and $\phi$ is the dilaton
\jsch. In
this way, M-theory geometrizes the conjectured $ {\rm SL}(2, {\bf
\rm Z})$
symmetry of type IIB string theory. At this stage, the type IIB string is
compactified on a circle of finite circumference. As we take the
area of the
torus to zero, the circle will decompactify. In this limit, we
expect to see a
new dimension grow in M-theory. Without loss of generality, we
shall assume the
torus is rectangular, and take the two
transverse
coordinates $X^1,X^2$ to be compactified on circles of
circumference $L_1,L_2$:
$$
\eqalign{0&<X^1<L_1 \cr
0&<X^2<L_2. \cr}
$$
To keep the IIB coupling fixed, we will hold the ratio $ { L_1 / L_2} $
constant, and consider the limit,

\eqn\1{\eqalign{L_1& \to 0 \cr
L_2& \to 0, \cr
}}
which should yield ten-dimensional type IIB theory. The extra spatial
coordinate
conjugate to
the two-brane
wrapping number will be called $Y$. Thus the eight transverse spatial
coordinates of IIB theory are $(X^3, \ldots, X^9,Y)$.

Before taking the limit \1, the quantum of energy associated with a
wrapped two-brane is given by,

\eqn\2{E_{wrap}={ L_1L_2\over (2\pi)^2 (l_{11}^p)^3},}
where $l_{11}^p$ is the eleven-dimensional planck length. If we identify
this wrapping energy with
the energy of the first Kaluza-Klein excitation of a massless
particle, then
the compactification circumference of the $Y$ coordinate is

\eqn\3{L_Y={2\pi \over E_{wrap}}={(l_{11}^p)^3 (2\pi)^3 \over L_1L_2}.}
We will eventually take the limit \1\ in which $L_Y \to \infty$,
but before
doing so let us compactify one additional dimension $X^3$. Therefore, our
starting point is M-theory on a three-torus.

Now, let us consider the M(atrix) formulation of toroidally compactified
M-theory. One way of describing compactifications of M(atrix)
theory is to
begin
with zero-branes on a d-dimensional space. Recall that the dynamics of
zero-branes is governed by open strings connecting the various
zero-branes. If
the compactification space has a nontrivial fundamental group, we
also have to
take into consideration strings that wind around the various
nontrivial cycles
in the space. Nice compactifications will preserve a large degree of
supersymmetry, and in these cases, we expect to be able to neglect higher
string
modes, and restrict our attention only to the massless excitations
of the open
strings. A toroidal compactification is certainly a nice choice since the
amount
of supersymmetry preserved by the compactification is maximal. So,
we begin by
considering zero-branes on a small d-dimensional torus. It is easy
to see,
either by T-duality or explicit construction \wt, that the relevant
dynamics is
nicely described by $U(N)$ supersymmetric Yang-Mills on the dual
torus. We
shall
eventually take $N \rightarrow\infty,$ although the following
argument is valid
for finite $N$.  As we make our original torus smaller, we will be
probing the
infra-red dynamics of the d+1-dimensional Yang-Mills theory. In
this respect,
compactification on the three-torus is very special since the model is
conformally invariant. For $d<3$, the Yang-Mills theory becomes strongly
coupled
in the infra-red, while for $d>3$, the theory is free in the infra-red.

This way of describing M(atrix) compactifications on the three-torus was
studied
in \tdsd\ using the properties of 3+1-dimensional Yang-Mills with four
supersymmetries. The parameters of the Yang-Mills theory were
derived in \tdsd,
and in particular, the dimensionless Yang-Mills coupling constant
is given by,

\eqn\4{g^2={(2\pi)^4 (l_{11}^p)^3\over L_1L_2L_3}.}
In this formulation, any symmetry between $L_Y$ and $L_3$ is
hidden. To exhibit
a symmetry, we shall choose the two lengths $L_Y$ and $L_3$ to be
equal to one
another.  Since
$Y$ and
$X^3$ are compactified on identical circles, the rotation symmetry in the
$Y,X^3$ plane is, for the moment,
broken to a discrete subgroup generated by rotations by multiples
of $\pi/2$.
We
will demonstrate this symmetry in the following way.

 From \3, and the equality of $L_Y$ and $L_3$, we find

\eqn\5{L_1L_2L_3= (2\pi)^3 (l_{11}^p)^3.}
Combining this relation with \4, we obtain:

\eqn\6{g^2=2\pi.}
Now this is a very special value of the coupling. Recall that
S-duality implies
that couplings related by,

\eqn\7{{\tilde g} = {2\pi \over g},}
describe identical theories related by electric-magnetic
interchange. Eq \6\
then implies that we are working at the self-dual point at which
the theory is
invariant under a rotation of electric charge into magnetic charge.
In turn,
this suggests
that the hidden spatial rotation invariance is nothing more than the
electric-magnetic self-duality.

To see the connection between $Y, X^3$ rotations and
electric-magnetic rotations, let us consider the conjugate momenta
$P_Y$ and
$P_3$. As we have
mentioned, $P_Y$
is proportional to the wrapping number of membranes wrapped  on
the $X^1-X^2$
torus. Let us return momentarily to M-theory on this two-torus. The
wrapping
number of the membranes is easily identified with a flux in the M(atrix)
formulation of M-theory on a two-torus in the following way. Instead of
starting
with a theory of pure zero-branes on the two-torus, we can consider
a theory of
$N$ zero-branes and some number of two-branes. T-dualizing both
cycles of the
torus exchanges the number of zero-branes and two-branes. The zero
brane charge
is proportional to,
$$ \int_{T^2} \tr F,$$
where $F$ is the two-form field strength for the $U(N)$ gauge
theory describing
the $N$ two-branes. Hence, this wrapping number is identified with the
abelian magnetic flux on this torus, as also mentioned in \rM. For
compactification on a
three-torus,
similar arguments given in \ganor, identify the number of membranes
wrapped on
the $1-2$ plane with the amount of abelian magnetic flux in the
$1-2$ plane. On
the other hand, the electric flux along the $X^3$ cycle was
identified with the
momentum conjugate to $X^3$ \tdsd. These two fluxes are rotated into one
another
by electric-magnetic duality!

Now consider the limit in which $L_3 \to \infty$. In this limit, the
rotational
invariance relating $X^3$ to the other transverse coordinates is
restored. Since we
have demonstrated a discrete symmetry relating $X^3$ and $Y$, we
must also have
continuous rotations between $Y$ and the remaining transverse
coordinates. In
fact, the entire
rotational
$O(8)$ symmetry then follows.

What does this argument imply about the resulting 2+1-dimensional
Yang-Mills
theory? Let us begin by considering the abelian 2+1-dimensional
theory that
describes a single two-brane of type IIA string theory wrapped on a
two-torus.
This theory has a manifest $O(7)$ symmetry rotating the seven
scalar fields
into
one another. In this case, it is easy to see that we can replace the
vector-field by a scalar parametrizing a compact direction using
vector-scalar
duality in three dimensions. This argument has been used in \rDL, \rtown\ and
\rschmid\ to understand the eleven-dimensional origin of the type IIA
two-brane.
As we shrink the area of the torus to zero, the compact direction
corresponding
to this dual scalar decompactifies, and we recover the full $O(8)$
symmetry.
Our
interest is actually with the non-abelian generalization
corresponding to $N$
two-branes wrapped on the two-torus. In this case, we cannot simply
dualize the
non-abelian vector-field. Nevertheless, from our preceeding
discussion, we know
that there should exist a superconformal theory in the infra-red
with global
symmetry enhanced from $O(7)$ to $O(8)$. It would interesting to
show that this
fixed point exists directly in the 2+1-dimensional Yang-Mills theory, and
indeed, Seiberg has argued for the existence of such a fixed point
directly in
the three dimensional theory \rNS. Furthermore, Banks and Seiberg
have also
studied type IIB strings in an alternate version of compactified
M(atrix)
theory and come to similar conclusions to our own \rBS.

In light of our previous discussion, we can now establish a
dictionary between
type IIB $(p,q)$ strings, and backgrounds in the 3+1-dimensional
Yang-Mills
theory. Note that in the type IIB theory, there are two compact
directions
corresponding to $X^3$ and $Y$. If we take a membrane wrapped
around the $X^1$
cycle to correspond to a fundamental type IIB string, then a
D-string will
correspond to a membrane wrapped around the $X^2$ cycle. Wrapping
the membrane
$p$ times around $X^1$, and $q$ times around $X^2$, which we will call a
$(p,q)$
cycle, will give a $(p,q)$ string. Wrapping the remaining leg of
the membrane
around the $X^3$ direction will correspond to a magnetic flux in
the plane
determined by $X^3$ and the $(p,q)$ cycle. On the type IIB side,
this membrane
configuration corresponds to wrapping a $(p,q)$ string around the $X^3$
direction. What corresponds to wrapping a $(p,q)$ string around the $Y$
direction? We simply need to apply electric-magnetic duality, which will
exchange the $X^3$ and $Y$ directions. Therefore a $(p,q)$ string
wound around
the $Y$ direction will be realized in the Yang-Mills theory as an
electric flux
along the direction orthogonal to the plane determined by $X^3$ and
the $(p,q)$
cycle. Configurations corresponding to higher dimensional branes in
type IIB
will correspond to nontrivial configurations of the scalar fields in the
Yang-Mills theory. Unlike the wrapped $(p,q)$ strings,
understanding these
configurations will require a study of the Yang-Mills theory in the
$N\rightarrow \infty$ limit.

\bigbreak\bigskip\bigskip\centerline{{\bf Acknowledgements}}\nobreak

We are very grateful to Nathan Seiberg for enlightening discussions
concerning
the existence of a 2+1 dimensional fixed point with O(8) symmetry.
It is also a
 pleasure to thank T. Banks, R. Gopakumar, S. Ramgoolam, S. Shenker
and E.
Witten for helpful discussions. The work of S.S. is supported by
NSF grant
DMS--9627351, while that of L.S. is supported by NSF grant phy-9219345A2.

\vfill\eject

\listrefs
\bye